\documentclass[conference]{IEEEtran}
\IEEEoverridecommandlockouts

\usepackage{cite}
\usepackage{amsmath,amssymb,amsfonts,amsthm}
\usepackage{graphicx}
\usepackage{textcomp}
\usepackage{xcolor}

\ifCLASSINFOpdf
\else
\fi
\usepackage{soul}
\usepackage{subfigure}
\usepackage{tabularx,booktabs}
\usepackage{url}
\usepackage{multirow}
\usepackage{boldline}
\usepackage{pifont}
\theoremstyle{definition}

\usepackage{algorithm}
\usepackage{algpseudocode}
\usepackage{mathtools}

\usepackage{subfiles}

\begin{document}

\title{On the Decentralization of Blockchain-enabled Asynchronous Federated Learning
\thanks{This work was funded by the IN CERCA grant from the Secretaria d'Universitats i Recerca del departament d'Empresa i Coneixement de la Generalitat de Catalunya, and partially by the Spanish project PID2020-113832RB-C22(ORIGIN)/MCIN/AEI/10.13039/50110001103, the grant CHIST-ERA-20-SICT-004 (SONATA) by PCI2021-122043-2A/AEI/10.13039/501100011033 and European Union’s Horizon 2020 research and innovation programme under Grant Agreement No. 953775 (GREENEDGE).}
}

\author{\IEEEauthorblockN{Francesc Wilhelmi}
\IEEEauthorblockA{CTTC/CERCA\\
Barcelona, Spain \\
fwilhelmi@cttc.cat}
\and
\IEEEauthorblockN{Elia Guerra}
\IEEEauthorblockA{CTTC/CERCA\\
Barcelona, Spain \\
eguerra@cttc.es}
\and
\IEEEauthorblockN{Paolo Dini}
\IEEEauthorblockA{CTTC/CERCA\\
Barcelona, Spain \\
paolo.dini@cttc.es}
}

\maketitle

\begin{abstract}
Federated learning (FL), thanks in part to the emergence of the edge computing paradigm, is expected to enable true real-time applications in production environments. However, its original dependence on a central server for orchestration raises several concerns in terms of security, privacy, and scalability. To solve some of these worries, blockchain technology is expected to bring decentralization, robustness, and enhanced trust to FL. The empowerment of FL through blockchain (also referred to as \emph{FLchain}), however, has some implications in terms of ledger inconsistencies and age of information (AoI), which are naturally inherited from the blockchain's fully decentralized operation. Such issues stem from the fact that, given the temporary ledger versions in the blockchain, FL devices may use different models for training, and that, given the asynchronicity of the FL operation, stale local updates (computed using outdated models) may be generated. In this paper, we shed light on the implications of the FLchain setting and study the effect that both the AoI and ledger inconsistencies have on the FL performance. To that end, we provide a faithful simulation tool that allows capturing the decentralized and asynchronous nature of the FLchain operation. 
\end{abstract}


\IEEEpeerreviewmaketitle

\section{Introduction}

Edge intelligence, by means of edge computing, is expected to bring artificial intelligence (AI) applications and optimizations closer to users by providing low-latency responses both in training and inference phases~\cite{zhou2019edge}. Moreover, the edge computing paradigm is expected to provide more efficient usage of resources for handling massive amounts of user-generated data with respect to traditional centralized learning approaches, typically hosted at large data centers.

One paradigm increasing in popularity to enable edge intelligence is federated learning (FL), whereby a set of participants exchange model parameters to build a collaborative model~\cite{konevcny2016federated}. FL has been typically realized through the orchestration of a central server, which is responsible for gathering local updates computed by FL clients, aggregating the information to generate a new version of the global model, and distributing new models to clients. Following this approach, an FL algorithm can potentially reduce the overheads and enhance the privacy of its centralized counterpart.

The traditional centralized FL approach, however, has serious disadvantages that prevent its adoption in many large-scale settings~\cite{li2020federated}. First, the central server represents a bottleneck and a single point of failure, which can potentially compromise performance and security. Second, given the devices' heterogeneity in terms of computation, communication, and availability, centralized FL entails issues related to fault tolerance (e.g., straggler effect). Moreover, the orchestration of FL devices depends on a single server, typically controlled by a single entity, which is detrimental to the democratization of the training procedure. In this regard, an arbitrary set of devices could be selected in each FL iteration to satisfy the interests of a single party, therefore leading to biased models (e.g., taking into consideration only portions of data that are relevant for the orchestrator). 

To address the issues of centralization in FL, we focus our attention on blockchain, which can enable a secure, reliable, and transparent server-less realization of FL, i.e., \textit{FLchain}~\cite{kim2019blockchained}. By means of cryptographic proof, blockchain provides trust to a federated ecosystem whereby multiple (often unreliable) parties can cooperate to train a collaborative global, each one providing insights into their own data. In FLchain, blockchain nodes (e.g., miners) maintain a distributed ledger that stores FL model updates in blocks that are generated at regular intervals. To secure the data and achieve consistency (i.e., nodes should agree on the same history of the ledger), miners run a mining mechanism like Proof-of-Work (PoW) and apply a certain set of consensus rules (e.g., the main chain is the one with more power invested).

The FLchain solution, on the one hand, allows removing the single-point-of-failure problem of centralized FL, thus solving scalability issues, and providing enhanced trust thanks to blockchain properties. On the other hand, the blockchain's decentralized architecture leads naturally to an asynchronous operation whereby local updates are shared as they become available, without any kind of coordination. Moreover, FL devices may use potentially different global models for training, as a result of the temporary inconsistencies in the ledger (for instance, due to forks). 

In this paper, we aim to shed light on the implications of the server-less FL operation when realized through blockchain technology. In particular, we focus on the asynchronous nature of the learning procedure and on the ledger inconsistencies that stem from the blockchain operation, whose impact remains unclear. More specifically, we delve into the effects of using stale and disparate model updates in FL training. In this regard, we feature an age of information (AoI)-based metric~\cite{yates2021age}, introduced as age of block (AoB), to evaluate the freshness of model updates in FLchain. 

To drive our analysis, we provide an integration of the BlockSim simulator~\cite{alharby2019blocksim}, which characterizes any type of blockchain system, with FL operation through Pytorch~\cite{NEURIPS2019_9015}. We name this integration Block\textit{FL}Sim~\cite{wilhelmi2022flchainsim}, which, to the best of our knowledge, is the first of its kind. Our implementation is evaluated for a federated text recognition application, using the MNIST dataset~\cite{lecun1998gradient}.

The rest of the paper is structured as follows: Section~\ref{section:related_work} overviews the literature on server-less realizations of FL, including FLchain. Section~\ref{section:system_model} presents the details of the blockchain-enabled FL implementation, which deals with asynchronous model updates. Section~\ref{section:performance} provides a set of simulation results to showcase the performance of the proposed asynchronous FL application for text and image recognition tasks, as well as the set of implications that stem from the blockchain operation. Section~\ref{section:conclusions} concludes the paper.

\section{Related Work}
\label{section:related_work}



The trend of distributing and decentralizing ML operations has been embraced as an appealing solution for addressing many issues of centralization (connectivity, privacy, and security). One appealing distributed learning solution is FL~\cite{konevcny2016federated}, whereby different devices collaborate to train a model by exchanging model parameters that are obtained from local (and unshared) data. The FL framework has received a lot of attention in the recent years, and its implementation in real-world applications is extensive~\cite{aledhari2020federated}. Nevertheless, traditional FL settings still require the figure of a central server, responsible for clients orchestration and model aggregation.

A prominent solution to fully decentralize FL is FLchain~\cite{kim2019blockchained, majeed2019flchain, bao2019flchain}, where a blockchain system allows FL clients to submit and retrieve model updates without the need for a central server. Blockchain technology, beyond enabling decentralization, provides important enablers for trust, including security, privacy, and traceability, thus outperforming other existing decentralized FL solutions like the ones in~\cite{hu2019decentralized} (based on gossip learning) or \cite{roy2019braintorrent} (based on BitTorrent). FLchain, thanks to the abovementioned blockchain properties, opens the door to an unprecedented way of building powerful collaborative ML models through the participation of multiple untrusted parties, and without the need for any regulating third party. In this regard, we find FLchain realizations for novel use cases like autonomous vehicles~\cite{pokhrel2020federated} or fog computing~\cite{qu2020decentralized}. For further details on the integration of FL and blockchain, we refer the interested reader to the surveys in~\cite{nguyen2021federated, hou2021systematic}.

While promising decentralization and trust, FLchain entails a set of limitations that, to the date, have been barely studied. One important consequence of the architectural shift to decentralization stems from the way model updates are included into blockchain blocks. Unlike for centralized settings where a server decides the set of participating users in each FL round, in blockchain, FL clients submit model updates asynchronously. Moreover, these local updates are incorporated into blocks of different length, which depends on blockchain configuration parameters like the maximum block size and the block interval. An important aspect of asynchronous learning stems from the fact that FL devices may contribute to global training with outdated models (i.e., models that have been updated using past gradient information, or models that have not been updated at all). This poses a set of interesting challenges and trade-offs for incorporating data from slow devices. The asynchronous property of FLchain has been considered in a limited (and recent) number of contributions~\cite{lu2020blockchain, liu2021blockchain, wang2022asynchronous, feng2022bafl}.

The other important implication of FLchain is related to the temporal inconsistencies of the ledger across blockchain nodes, which results from the partitioning of the blockchain due to forks. Since blockchain nodes provide global models to FL devices, forking events may lead FL nodes to train on different models. This aspect has not been examined in the literature and, typically, blockchain systems have been assumed to act as a central server with a perfect synchronization among blockchain nodes (see, e.g.,~\cite{pokhrel2020decentralized}). In this paper, we focus on realistic implementations of FLchain, where FL devices carry out the asynchronous training operation in the presence of ledger inconsistencies. Moreover, we delve into the effect of incorporating or working with old models as a result of the asynchronous FL operation implicit in blockchain architecture. For that purpose, we study the staleness problem and analyze the AoI of the FL models generated by FL devices.

\section{Decentralized Aynchronous FLchain}
\label{section:system_model}

We adopt a PoW-based blockchain system, where transactions are organized in blocks (with size $S^B$), which are generated (mined) at regular intervals $BI$ (e.g., every 15 seconds)~\cite{gervais2016security}. In FLchain, the transactions are the local updates submitted by FL devices, which are gathered and spread throughout the blockchain's peer-to-peer (P2P) network by a set of full blockchain nodes $\mathcal{M}$ (with $M = |\mathcal{M}|$), acting as miners, as well. At this point, it is important to acknowledge that the blockchain mining procedure and the FL operation are executed in parallel, mostly independently one from the other.\footnote{Notice that blocks are mined sequentially one after another, regardless of the incoming FL model updates.}

\begin{figure}[ht!]
\centering
\includegraphics[width=\linewidth]{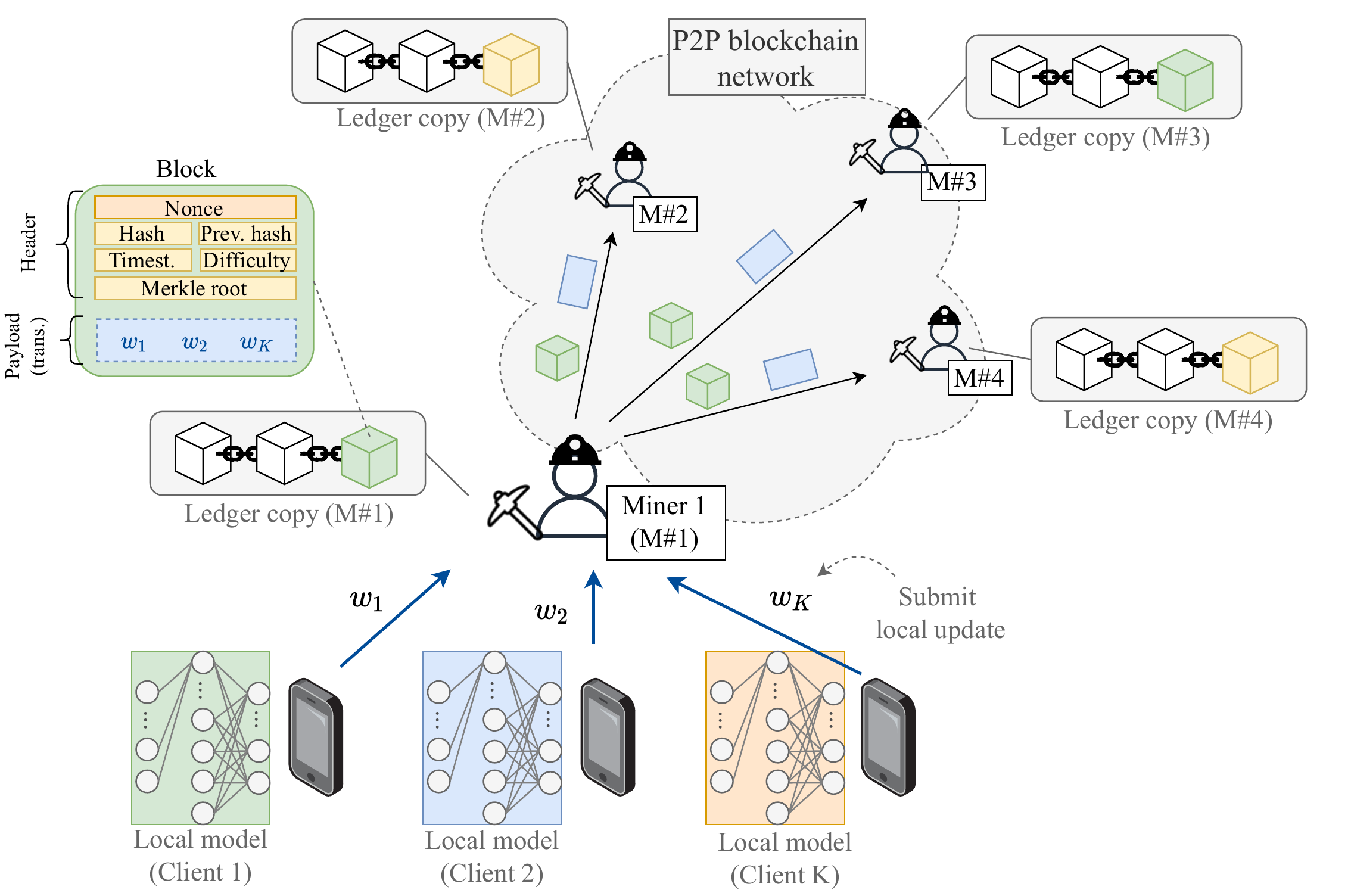}
\caption{Overview of the FLchain operation.}
\label{fig:blockchain_intro}
\end{figure}

When it comes to the FL operation (illustrated in Fig.~\ref{fig:blockchain_intro}), a dataset $\mathcal{D}$ distributed across a set of clients $\mathcal{K}$ (with $K = |\mathcal{K}|$) is used to train a global model $w$ collaboratively. Under the FLchain setting, each FL client $k \in \mathcal{K}$ with computational power $\xi^{(k)}$ and local dataset $\mathcal{D}^{(k)}$ of length $D^{(k)}$ generates local model updates $w^{(k)}$ by following the next asynchronous steps (and described also in Algorithm~\ref{alg:fl}):
\begin{enumerate}
    \item Initialization with global model $w_0$.
    \item Compute a local model update $w^{(k)}$, using the latest received global model. To do so, the client runs $E$ epochs of SGD based on the target local loss function $l^{(k)}$ and the batch size $B$ applied to $D^{(k)}$ local data points, so $w^{(k)} \leftarrow w^{(k)} - \eta \nabla l^{(k)}(w,\mathcal{D}^{(k)})$, where $\eta$ is the learning rate.
    \item Submit the local update $w^{(k)}$ to the blockchain, to be included in a future block by a given miner.
    \item Receive a new block and extract $U$ updates for global model aggregation. Aggregation is performed following FedAvg, so that $w \leftarrow \sum_{u=1}^{U} \alpha^{(u)} w^{(u)}$, where $\alpha^{(u)}$ is a scalar indicating the importance of weight $w^{(u)}$.
    \item Repeat steps (2)-(4) until convergence.
\end{enumerate}

\begin{algorithm}[h!]
	\caption{Dec. Asynchronous FLchain (a-FLchain)}\label{alg:fl}
	\begin{algorithmic}[1]
	    \State \textbf{Initialize:} $\eta$, $E$, $B$, $\mathcal{D}^{(k)}$, $\xi^{(k)}$
	    \Procedure{ModelTraining():}{}
		\While{training}
			\State Retrieve block $b$ (with $U$ updates) from miner $m$
			\State Model aggregation: $w^{(k)} \leftarrow \sum_{u=1}^{U} \alpha^{(u)} w^{(u)}$
    		\For{$e=1,\ldots,E$}
        		\State $w^{(k)}_{e}=w_{e}^{(k)}-\eta \nabla l^{(k)}_{e}(w,\mathcal{D}^{(k)},B,\xi^{(k)})$
    		\EndFor
    		\State Send local model update $w_{E}^{(k)}$ to miner $m$
		\EndWhile
		\EndProcedure
	\end{algorithmic}
\end{algorithm}

Different from typical centralized FL settings, in FLchain, each client $k\in \mathcal{K}$ maintains its own FL timeline according to the set of blocks it uses for keeping track of global model updates. This is due to the decentralized operation of the blockchain, whereby blockchain nodes (miners) may use inconsistent versions of the ledger. In particular, when a miner generates a block of depth $n$ (the genesis block has a depth equal to $0$) as a result of the mining procedure (e.g., solving a computational-intensive puzzle), it propagates it through the blockchain P2P network for validation. A block is validated when the rest of the miners append it to their version of the ledger and start working on the next block (with depth $n+1$). The version of the ledger accepted by the majority of the miners is referred to as the main chain. However, since any participating miner can generate blocks, different miners may come across a valid solution for appending a block with the same depth simultaneously, which would lead to forks (i.e., divergent (inconsistent) histories of the ledger). These inconsistencies are eventually solved thanks to consensus, which states that the longest chain (i.e., the one with the highest invested computational power) is the valid one. The longest chain is also referred to as main chain in the rest of paper. Accordingly, as the main chain increases, miners working on forked chains will abandon their chains and switch to the valid one. In the literature (see, e.g.,~\cite{feng2022bafl, liu2021blockchain}), a widely adopted assumption is that all the FL clients in FLchain can perfectly access the blockchain's main chain simultaneously, thus disregarding the effect of temporal ledger inconsistencies.


Regarding the freshness of global model updates in FLchain, we define the age of block (AoB) as the mean peak AoI~\cite{yates2021age} across all the transactions in a block. The AoB can be computed as the mean difference between the time clients generate local updates and the time it takes to include them into a block. Formally, the AoB of block $i$ ($\Delta_i$) is given by: 
\begin{equation}
    \Delta_i = \frac{1}{U}\sum_{j=1}^{U}\delta_{i,j} = \frac{1}{U}\sum_{j=1}^{U} t_{i,j}^{(2)} - t_{i,j}^{(1)},
\end{equation}
where $U$ is the number of client updates included in block~$i$, $t_{i,j}^{(1)}$ is the time at which client $j$ generates a local model update, and $t_{i,j}^{(2)}$ is the time at which node~$j$'s local update is included into block $i$. It is worth noting that the AoB depends on FL clients' communication capabilities and blockchain parameters (e.g., block size, block interval, capacity of P2P links).


\section{Performance Evaluation}
\label{section:performance}

We study the impact of the the introduced AoB metric on the learning performance under  different types of blockchain networks (e.g., Ethereum) and blockchain parameters (e.g., the maximum block size). The targeted FL application is based on hand-written digits recognition. For that, we define a four-layered feed-forward neural network (FNN) model (with 784, 200, 200, and 10 neurons in each layer), which is trained individually by FL clients using SGD on the well-known MNIST dataset~\cite{lecun1998gradient}, whose weights are aggregated using FedAvg~\cite{mcmahan2017communication}. The MNIST dataset consists of 60,000 and 10,000 data samples for training and test, respectively. For evaluation purposes, we have further split the test dataset into test (30\%) and validation (70\%).

All the simulations are done in Block\textit{FL}sim~\cite{wilhelmi2022flchainsim}, which allows studying in detail the real phenomena in FLchain. Table~\ref{tab:simulation_parameters} collects all the simulation parameters.\footnote{For the sake of openness and reproducibility, all the source code used in this paper is open and can be accessed at \url{https://gitlab.cttc.es/supercom/blockFLsim/-/tree/BlockFLsim} (commit: 8fa5c48e).} During the simulations, we keep track of \textit{(i)} the training accuracy of the main chain (i.e., for the FL devices including transactions in each block), \textit{(ii)} the training loss of every FL device when computing local model updates, and \textit{(iii)} the test accuracy of the resulting global model when the FL procedure stops.
\begin{table}[ht!]
\centering
\caption{Model/Simulation parameters}
\label{tab:simulation_parameters}
\begin{tabular}{@{}ccc@{}}
\toprule
Parameter & Description & Value \\ \midrule
$S^B$ & Max. block size & \{5, 10, 20\} trans.\\
$T_l$ & Transaction length & 796.84 kbits \\
$T_h$ & Block header length & 20 kbits \\ 
$M$ & Number of miners & 10 \\
$C_{p2p}$ & P2P links' capacity & \{10, $\infty$\} Mbps \\
$C_{n}$ & FL devices links' capacity & 1 Mbps \\
$BI$ & Block interval & \{5, 15, 60\} s \\
$NB$ & Number of simulated blocks & 50 \\
\hline
$N$ & Number of FL devices & \{10, 50, 100\} \\
$\xi$ & Devices comp. power & \{4.744, 83.000\} MIPS
\\
$E$ & Number of local epochs & 3 \\
$B$ & Batch size & 20 \\
$n^{(l)}_{fnn}$ & FNN's neurons per layer & [784, 200, 200, 10] \\
$a$ & Activation functions & ReLU / Softmax \\
$o$ & Optimizer & SGD \\
$\eta$ & Learning rate & 0.01 \\
$t_{fl}$ & Time to compute local updates & Exp($\xi$) \\
\bottomrule
\end{tabular}
\end{table}

\subsection{Performance in the main chain}

We evaluate the performance of FLchain in the main chain, which would match clients' performance if the blockchain's information could be simultaneously accessed by all the FL nodes across the network (i.e., without experiencing ledger inconsistencies). This assumption has been widely adopted in the literature but, as shown in the next subsections, does not hold in reality due to the network heterogeneity and temporal forks in the blockchain. Nevertheless, the main chain's accuracy is a solid indicator of the overall performance.

\begin{figure}[ht!]
\centering
\subfigure[Effect of $BI$]{\label{fig:main_chain_accuracy_a}\includegraphics[width=0.48\columnwidth]{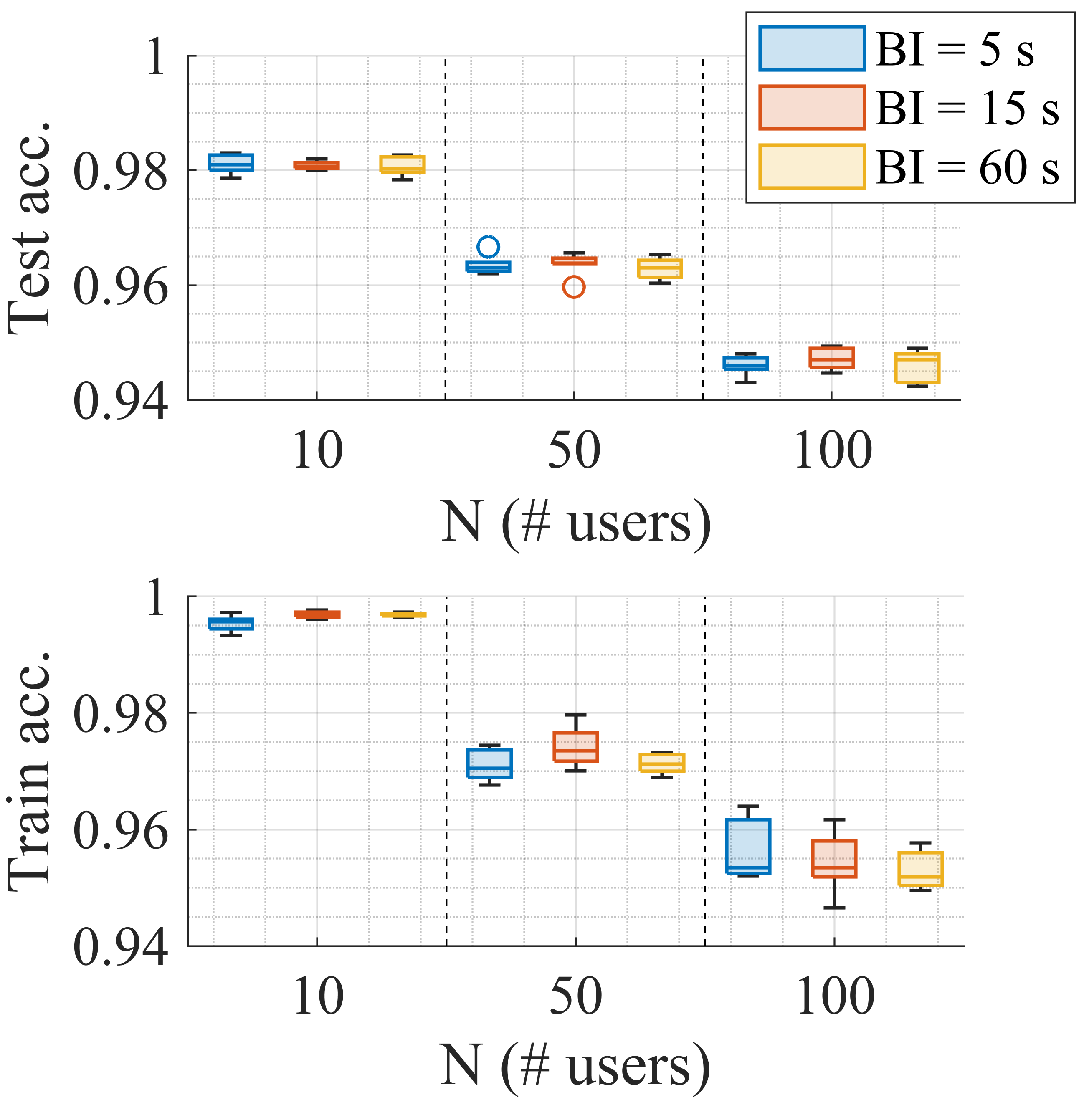}} 
\subfigure[Effect of $\xi$]{\label{fig:main_chain_accuracy_b}\includegraphics[width=0.48\columnwidth]{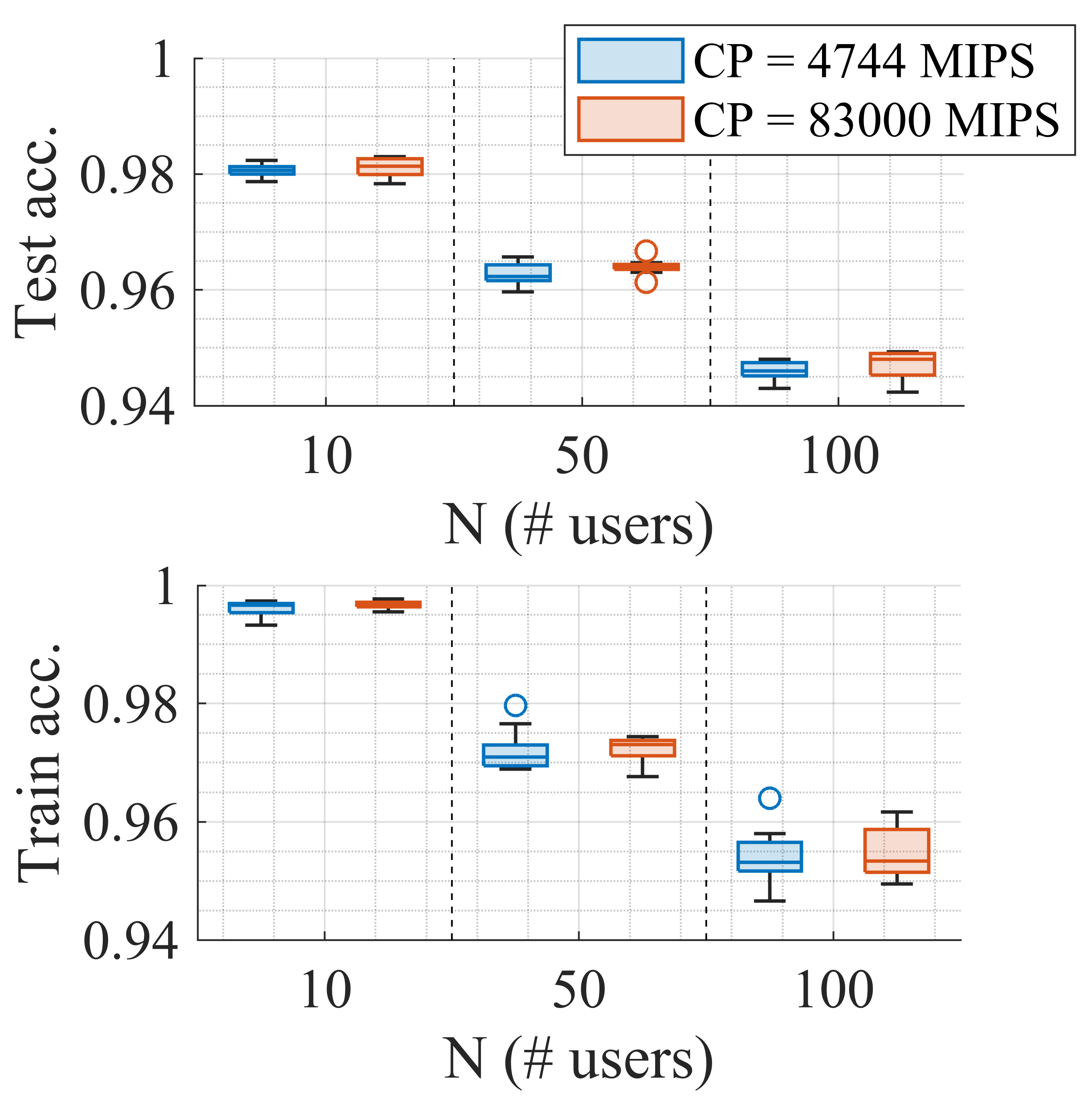}} 
\caption{Main chain's test and training accuracy in different scenarios}
\label{fig:main_chain_accuracy}
\end{figure}

Fig.~\ref{fig:main_chain_accuracy} shows the main chain's test and training accuracy for various scenarios. While Fig.~\ref{fig:main_chain_accuracy_a} focuses on the impact of the block interval $BI$ on the learning accuracy, Fig.~\ref{fig:main_chain_accuracy_b} illustrates the effect of the FL devices' computational capabilities. In particular, we consider FL scenarios of $N=\{10,50,100\}$ clients with two different types of computational power $\xi=\{$4.744, 83.000$\}$ MIPS (matching Raspberry Pi 2 and Intel Core i5-2500K computational capabilities). Furthermore, we depict different types of blockchain networks with $BI=\{5,15,60\}$ seconds and $S^B=\{5,10,20\}$ transactions. Notice that the impact of $S^B$ is captured in each box of the plot. The propagation delays in the blockchain are neglected in this part of the analysis (i.e., $C_{p2p} \approx \infty$).

As shown in Fig.~\ref{fig:main_chain_accuracy}, the test and training accuracy decreases as the number of clients increases. This is because the whole dataset is split into the considered number of devices, so FL local updates become more meaningful as $N$ decreases. Regarding the effect of $BI$ on the accuracy (see Fig.~\ref{fig:main_chain_accuracy_a}), it has a different effect depending on the selected number of FL devices. For small $N$ values, the block interval $BI$ has a moderate effect on the test and training accuracy. However, for $N=100$, we observe a decrease in the training accuracy as $BI$ increases, which is due to the extra delay required to incorporate all the clients' local updates in the global model. When it comes to the implications of using devices with different computational capabilities (see Fig.~\ref{fig:main_chain_accuracy_b}), we observe that the best results are achieved for $\xi=$83.000 MIPS.

\subsection{Age of Blocks}

\begin{figure}[ht!]
\centering
\includegraphics[width=\linewidth]{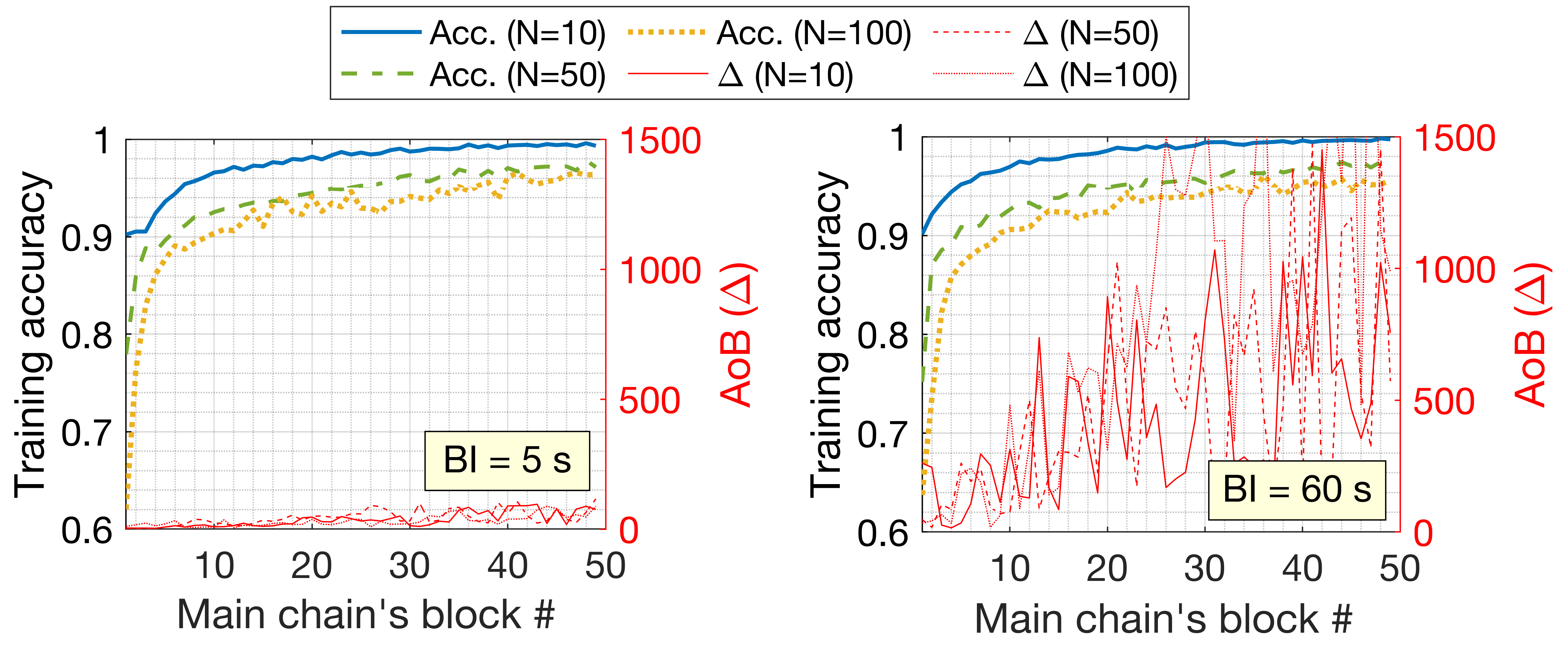}
\caption{Evolution of the AoB ($\Delta$) and the training accuracy for each main chain's block.}
\label{fig:impact_aoi_main_chain}
\end{figure}

We now focus on the impact of the AoB (presented in Section~\ref{section:system_model}) on the FL performance. Fig.~\ref{fig:impact_aoi_main_chain} illustrates, for each number of FL devices $N=\{10,50,100\}$, both the AoB and training accuracy achieved as the main chain increases. For the sake of illustration, we show the simulations for $BI=5$s and $BI=60$s, since they represent extreme cases with low and high AoB.

As it can be observed in Fig.~\ref{fig:impact_aoi_main_chain}, the training accuracy increases with the main chain's block number in all the cases. As previously observed, a low $N$ grants higher accuracy values. As for the AoB, an important conclusion is that, even if being very high (e.g., as for $BI=60$s), it has a low impact on the training accuracy. This indicates that, for the targeted FL application, untimely local updates can contribute to improve the performance of a collaboratively trained model. Another observation is that a higher AoB may be reached for high computational capabilities ($\xi$) and high block intervals ($BI$) because the the clients are processing faster than the miners and the blockchain system is not able to incorporate all the local updates into blocks in a timely manner.

\begin{figure}[ht!]
\centering
\includegraphics[width=\linewidth]{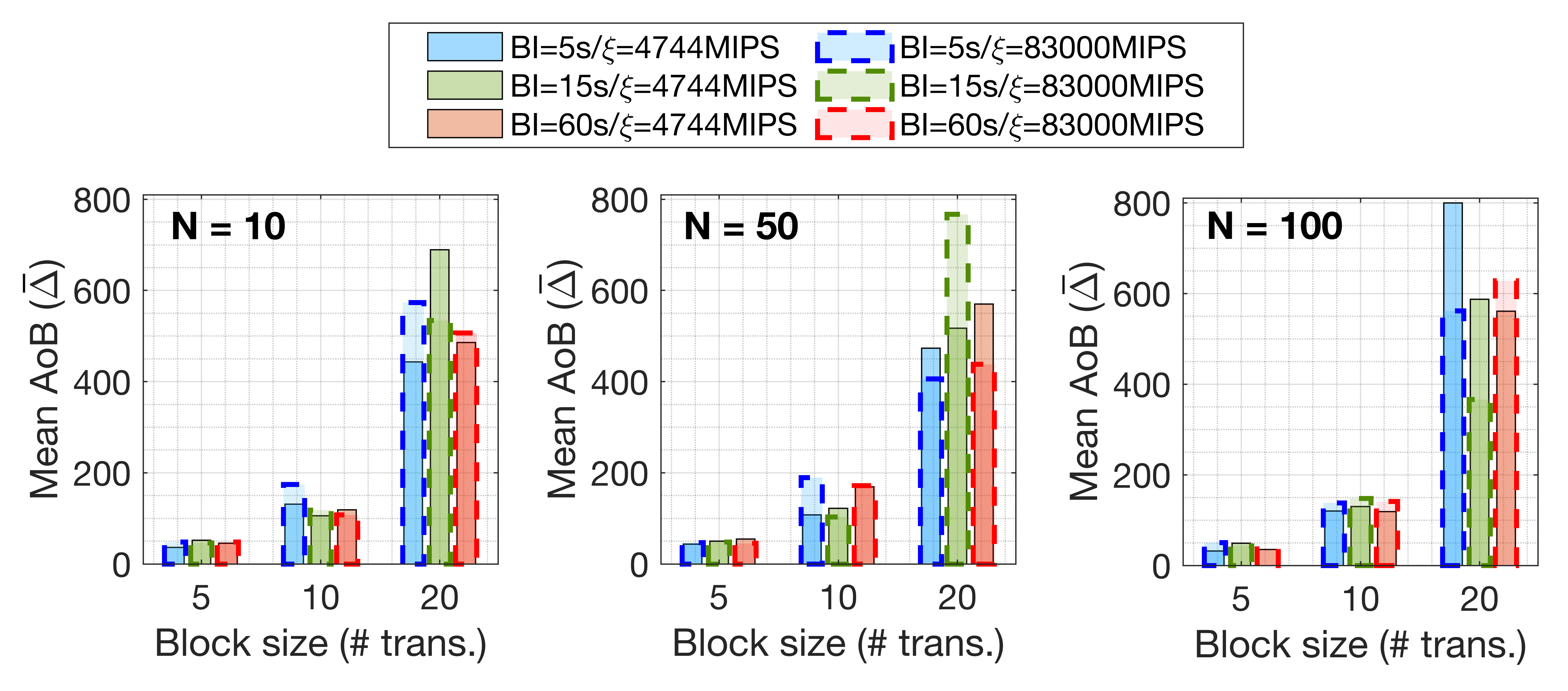}
\caption{Mean AoB ($\overline{\Delta}$) with respect to blockchain and FL parameters.}
\label{fig:block_age}
\end{figure}

Next, Fig.~\ref{fig:block_age} serves to further analyze the sensitivity of the AoB to blockchain parameters. Each subplot represents a different number of FL clients, whereas solid and dashed bars refer to the two types of FL devices in terms of computational capabilities ($\xi=$4.744 MIPS and $\xi=$83.000 MIPS, respectively). In this occasion, we plot the mean AoB ($\overline{\Delta}$) obtained throughout the entire simulations.

As previously illustrated in Fig.~\ref{fig:impact_aoi_main_chain}, Fig.~\ref{fig:block_age} shows that the mean AoB increases for larger numbers of FL devices ($N$). This effect is further accentuated for $\xi=$ 83.000. An important conclusion is that, due to the non-linearity shown by the AoB across different scenarios, blockchain parameters (e.g., block size, block interval) must be carefully selected to optimize the blockchain delay, which would allow minimizing the AoB. The optimization of blockchain parameters has been previously targeted in~\cite{wilhelmi2022end}.

\subsection{Blockchain inconsistencies}

An important consideration of FLchain lies in the models used by individual FL devices for training. These models are retrieved from the closest miner latest block, which, due to the blockchain's decentralized operation, may not be the same for all the FL clients. One remarkable reason for such type of inconsistencies lies in forks, which occur when two or more miners generate a valid block simultaneously. The acceptance of inconsistent blocks leads to forks, which create different temporary versions of the ledger. These types of forks are eventually solved thanks to the consensus protocol that states that the longest chain (the chain with the highest invested computational power) is the valid one. Nevertheless, during the FL operation, different FL devices can potentially work with different models in case forks occur.

\begin{figure}[ht!]
\centering
\subfigure[$S^B=5$ transactions]{\label{fig:training_accuracy_capacity_bi5_bs5}\includegraphics[width=.7\columnwidth]{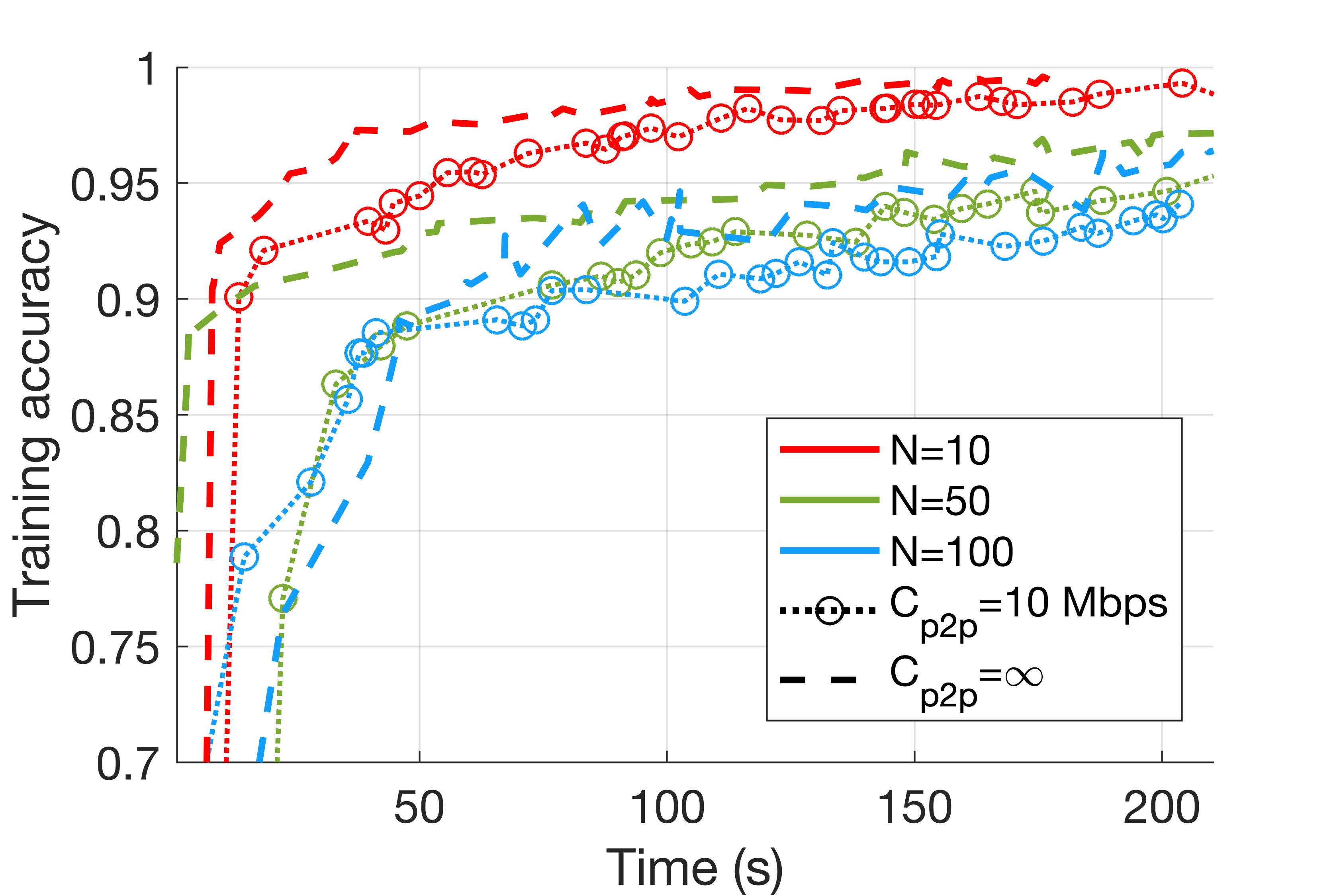}} 
\subfigure[$S^B=20$ transactions]{\label{fig:training_accuracy_capacity_bi5_bs20}\includegraphics[width=.7\columnwidth]{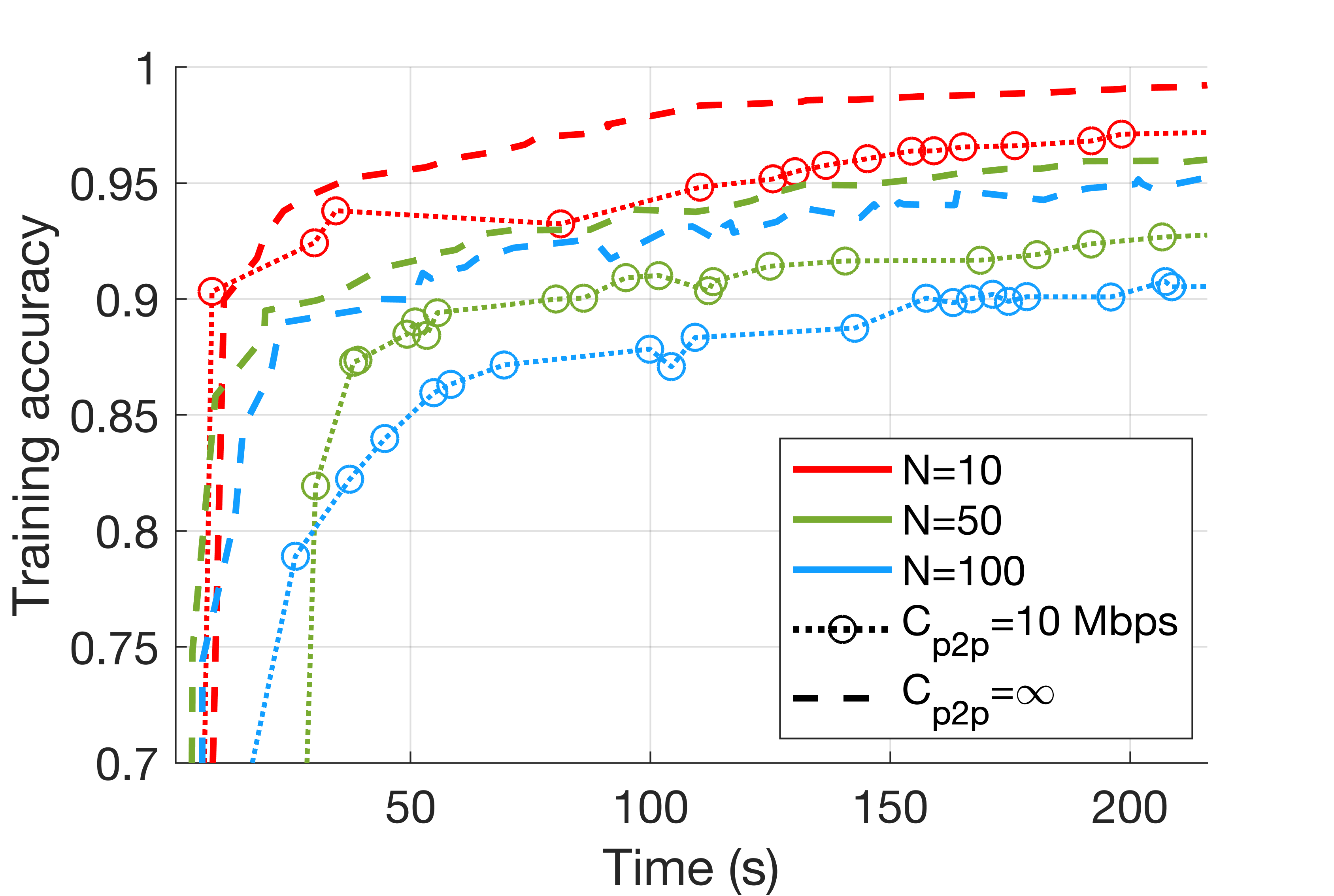}} 
\caption{Training accuracy for $C_{p2p} = 10$ Mbps and $C_{p2p} = \infty$ and $S^B = 5$ tr. and $S^B = 20$ tr.. }
\label{fig:training_accuracy_capacity_bi5}
\end{figure}

To showcase the effect that ledger inconsistencies in the form of forks have on the FL operation, we focus on the P2P links' capacity. In the previous subsections, we depicted the ideal case whereby blockchain delays were disregarded, so that all miners were virtually synchronized. This approach has been widely adopted in the existing literature, but differs from reality. To illustrate the effect of the capacity of the FL operation, Fig.~\ref{fig:training_accuracy_capacity_bi5} compares the FL training accuracy achieved during the simulation in two cases: \textit{i)} the blockchain communication delays are disregarded (i.e., for $C_{p2p}=\infty$), and \textit{ii)} forks can occur as a result of using a limited capacity in P2P links (i.e., for $C_{p2p} = 10$ Mbps). The block interval is fixed to $BI=5$s (the worst case with respect to forks), while $S^B=\{5,20\}$ transactions are considered.

As shown in Fig.~\ref{fig:training_accuracy_capacity_bi5}, the ideal case ($C_{p2p}=\infty$) provides better training accuracy than the realistic one ($C_{p2p}=10$ Mbps). Such a difference is more noticeable for $S^B=20$ transactions (see Fig.~\ref{fig:training_accuracy_capacity_bi5_bs20}), which, as shown later in this section, leads to a higher probability of experiencing forks. In particular, the stability of the blockchain favors the FL operation by providing timely and consistent models to FL devices. In contrast, for $C_{p2p}=10$ Mbps, the effect of forks and model inconsistencies has a detrimental effect on the training procedure, which is slowed down and experiences more instability.

\begin{figure}[ht!]
\centering
\includegraphics[width=.8\columnwidth]{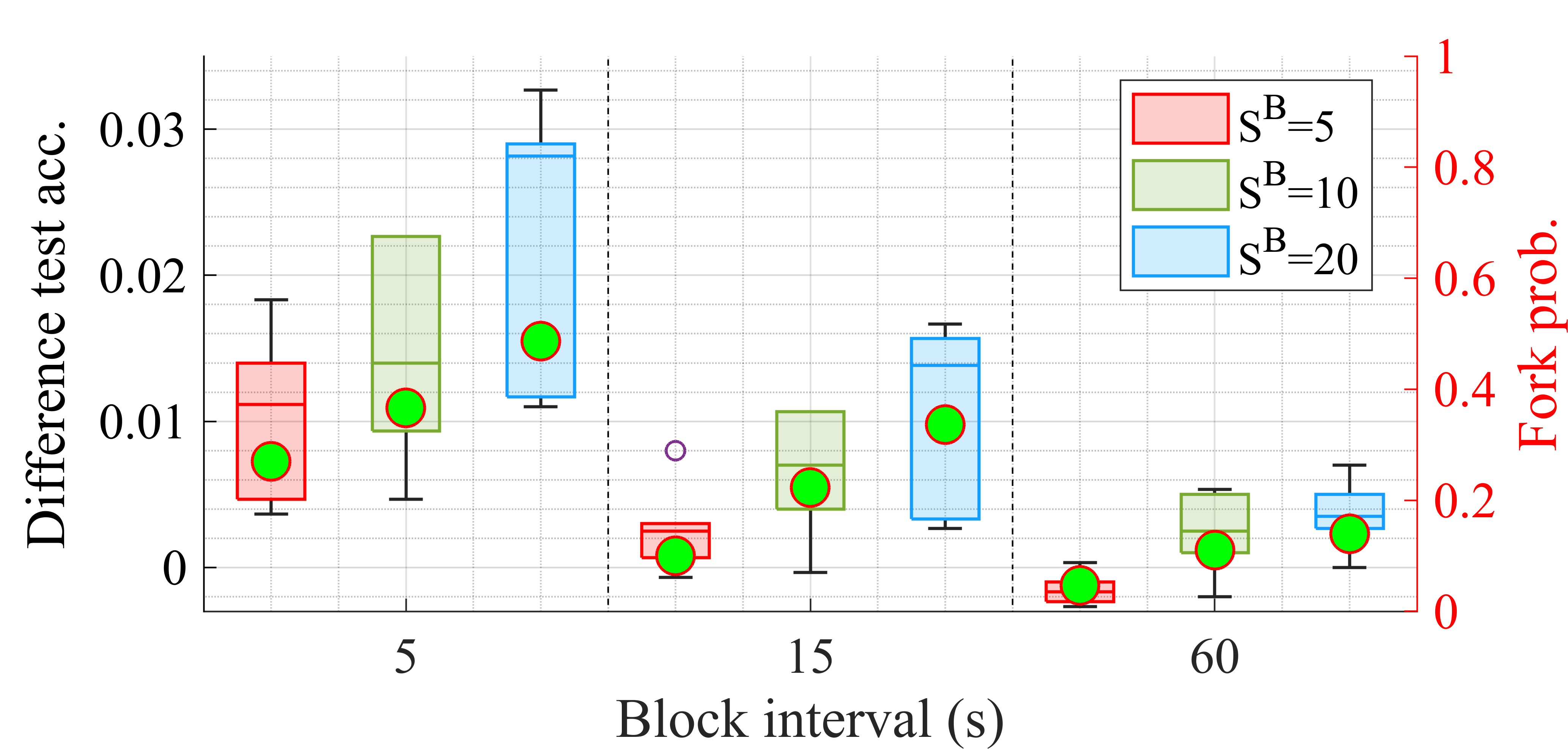}
\caption{FL test accuracy difference for $C_{p2p} = \infty$ and $C_{p2p} = 10$ Mbps. The fork probability is illustrated with green circles.}
\label{fig:impact_forks_main_chain}
\end{figure}

Finally, we focus on the impact that forks have on the test accuracy, which is obtained by evaluating the global model from the last simulated block on the test dataset. Fig.~\ref{fig:impact_forks_main_chain} shows the difference in the test accuracy for the two capacities considered above ($C_{p2p}=\{10~\text{Mbps}, \infty\}$). The results are shown for $BI=\{5,15,60\}$ s and $S^B=\{5,10,20\}$ transactions, and each boxplot includes the results for all the considered $\xi$ and $N$ values.

As shown in Fig.~\ref{fig:impact_forks_main_chain}, the forks occurrence increases with the maximum block size ($S^B$) and for shorter $BI$ values, which allows reducing the block propagation delay directly (the most stable configuration is obtained for $BI=60$ s and $S^B=5$). As for the difference in the test accuracy, it increases with the fork probability, where more significant ledger inconsistencies are expected. Still, the difference is very low, which suggests a very interesting result: even in the presence of inconsistencies, FL devices can still learn collaboratively.


\section{Conclusions}
\label{section:conclusions}

FLchain emerges as an appealing solution for secure and robust decentralized learning framework. However, the implications that decentralization of blockchain technology have on FL optimization have not been studied closely yet. In this paper, we have studied the implications of running an FL application in a blockchain, including the temporal inconsistencies and the age of information that are implicit in such a technology. To do so, we have proposed a simulation tool (Block\textit{FL}sim, an extension of BlockSim) for realistic FLchain operation. Our results show that both the age of information and model inconsistencies (forks) have a low impact on the accuracy of an FL application (trained with the MNIST dataset) running over a blockchain. This result suggests that FL is a robust framework that works effectively even if clients use outdated information or different models for collaborative training. Future work includes studying more complex datasets with unbalanced settings and non-IID data.

\ifCLASSOPTIONcaptionsoff
\newpage
\fi

\bibliographystyle{IEEEtran}
\bibliography{IEEEabrv,bibliography}

\end{document}